\documentclass[%
 reprint,
 amsmath,amssymb,
 aps,
 pra,
 floatfix,
 longbibliography,
]{revtex4-1}

\newcommand{\ket}[1]{|#1\rangle}
\newcommand{\bra}[1]{\langle#1|}

\newcommand{\XT}{X_\Phi}
\newcommand{\Phidc}{\bar{\Phi}}
\newcommand{\Phiac}{\tilde{\Phi}}

\usepackage{graphicx}
\usepackage[urlcolor=blue, hyperindex, colorlinks, bookmarks=true]{hyperref}
\usepackage[utf8]{inputenc}

\usepackage{microtype}

\begin{document}

\title{Methods for Measuring Magnetic Flux Crosstalk Between Tunable Transmons}

\author{Deanna M. Abrams}\email[Corresponding author: ]{deanna@rigetti.com}
\author{Nicolas Didier}
\author{Shane A. Caldwell}
\author{Blake R. Johnson}
\author{Colm A. Ryan}
\affiliation{Rigetti Computing, 2919 Seventh Street, Berkeley, California 94710}

\date{\today}

\begin{abstract}

In the gate model of quantum computing, a program is typically decomposed into a sequence of 1- and 2-qubit gates that are realized as control pulses acting on the system. A key requirement for a scalable control system is that the qubits are addressable --- that control pulses act only on the targeted qubits.  The presence of control crosstalk makes this addressability requirement difficult to meet. In order to provide metrics that can drive requirements for decreasing crosstalk, we present three measurements that directly quantify the DC and AC flux crosstalk present between tunable transmons, with sensitivities as fine as $0.001\%$. We develop the theory to connect AC flux crosstalk measures to the infidelity of a parametrically activated two-qubit gate. We employ quantum process tomography in the presence of crosstalk to provide an empirical study of the effects of crosstalk on two-qubit gate fidelity.

\end{abstract}

\maketitle

\section{Introduction}

As systems with many tens of qubits become available, building the associated control system remains a formidable technological integration challenge. A crucial requirement for the scalability of the control system is the addressability of the qubits --- that is, each control system action should affect only the intended target qubit(s), with no undesired crosstalk. While single qubit gate error rates are now routinely quoted in the $0.1\%$ range, and entangling gate error rates are at or below $1\%$~\cite{Hong:2019, KraglundAnderson:2019, ibmq:2018, Barends:2014, PhysRevA.93.060302, Barends2019, supremecy2019}, the reported gate fidelities are often given as the fidelity of isolated operations, which cannot capture the deleterious effects of any given gate on neighboring qubits from lack of addressability. Recent work has focused on how to characterize simultaneous operations as more qubits are operated in concert~\cite{Figgatt:2018, Rudinger:2018, Zhang:2018, Proctor:2018, Song:2017}. Many of these studies employ measurements such as simultaneous randomized benchmarking to measure the effects of performing qubit operations on many qubits in concert, and have noted degradation in computational performance as more qubits are operated in parallel~\cite{PhysRevLett.109.240504, PhysRevLett.122.200502, 2019arXiv190208543E, supremecy2019}. However, these high level measurements typically report fidelities, which combine many possible errors. In order to drive device level improvements, it is important to build a toolbox of quantitative measures that directly measure on-chip crosstalk, and which can then be used to predict the performance of higher level fidelity metrics. 

The coherent nature of crosstalk errors makes them particularly pernicious, as they can potentially add up over a train of applied gates in a given circuit~\cite{2019arXiv190705950G, Sanders_2015}. If crosstalk causes a control pulse to also affect unintended spectator qubits, then when scheduling gates, these side effects must be considered. Instead of the desired 1- or 2-qubit gate, we have a many-qubit gate, which, without measures to counter the crosstalk, prevents what could otherwise be simultaneous gates.  One mitigation approach is to use optimal control pulses that are robust to interference from crosstalk \cite{PhysRevB.97.045431}. Another approach is to characterize the crosstalk and then actively cancel it with compensating drives on the spectator qubits \cite{PhysRevA.93.060302, neill2018blueprint}. However, using compensating drives means that pulses on one channel become contextual --- dependent on what is being played simultaneously on other qubits --- and so requires many more variants of the pulse to be calibrated and configured in the control hardware. With either approach, the cost of mitigation is closely associated with the scale of the crosstalk. In optimistic scenarios of localized cross-talk, mitigation strategies may be tractable; however, rather than attempting to calibrate away crosstalk, we turn instead to providing accurate measures of crosstalk in order to inform device level changes that will obviate the need for mitigation.

Superconducting qubits are typically controlled by radio-frequency or microwave pulses that are calibrated with a specific frequency, duration, and amplitude in order to enact the desired quantum gates on specific qubits. For superconducting qubit architectures that utilize magnetic flux to tune the frequency of qubits, the current applied to generate flux through any given qubit's SQUID loop may also cause flux to thread other qubits' SQUID loops~\cite{PhysRevA.76.042319}, causing control crosstalk. The unimpeded flow of supercurrents along the ground plane of superconducting qubit chips means that the control fields used to operate qubits may have long range effects, coupling to unaddressed qubits in non-local and counter-intuitive ways.

We will define crosstalk as a ratio $\XT = d\Phi_A/d\Phi_B$; that is, given an \textit{adversarial} flux applied to qubit B, we want to measure the resulting flux felt by qubit A. By measuring crosstalk as a ratio, we define a metric that can be compared across qubits, devices, and architectures. Early demonstrations of flux-based entangling operations~\cite{dicarlo2009demonstration, reed2012realization, reed2013entanglement} and other flux-biased devices~\cite{PhysRevB.72.060506} came with reports of DC flux crosstalk in excess of 10\%, and later devices have shown DC flux crosstalk in the range of 0.1-10\%.~\cite{kelly2015state, riste2015detecting, neill2018blueprint, Kounalakis2018}. These crosstalk values were measured by various methods including: measuring the dependence of the qubit's frequency vs flux curve offset as a function of DC flux applied to other flux-controlled circuit elements~\cite{Kounalakis2018}; or tuning the relative amplitude of flux pulses on two control lines to cancel out any unintended phase accumulated on qubits due to crosstalk~\cite{neill2018blueprint}. The measured crosstalk matrix can then be orthogonalized in order to improve control addressability.

In this paper we present several direct methods for measuring both DC and AC flux crosstalk between two tunable transmons, and we employ quantum process tomography to investigate the effect of flux crosstalk on parametrically activated CZ gates. 
In section~\ref{sec:experimental-setup} we describe the experimental setup. In section~\ref{sec:dc-crosstalk}, we describe and implement two different methods for measuring flux crosstalk between two qubits at DC. In section~\ref{sec:ac-crosstalk}, we describe and implement a method for measuring flux crosstalk between two qubits as a result of AC pulses. In section~\ref{sec:cross-tomography}, we describe a method for directly measuring entangling gate errors incurred due to flux crosstalk, and show that simultaneous gate performance under the influence of crosstalk can be modeled theoretically.

\section{Experimental Setup}\label{sec:experimental-setup}
\begin{figure}
    \includegraphics[width=\columnwidth]{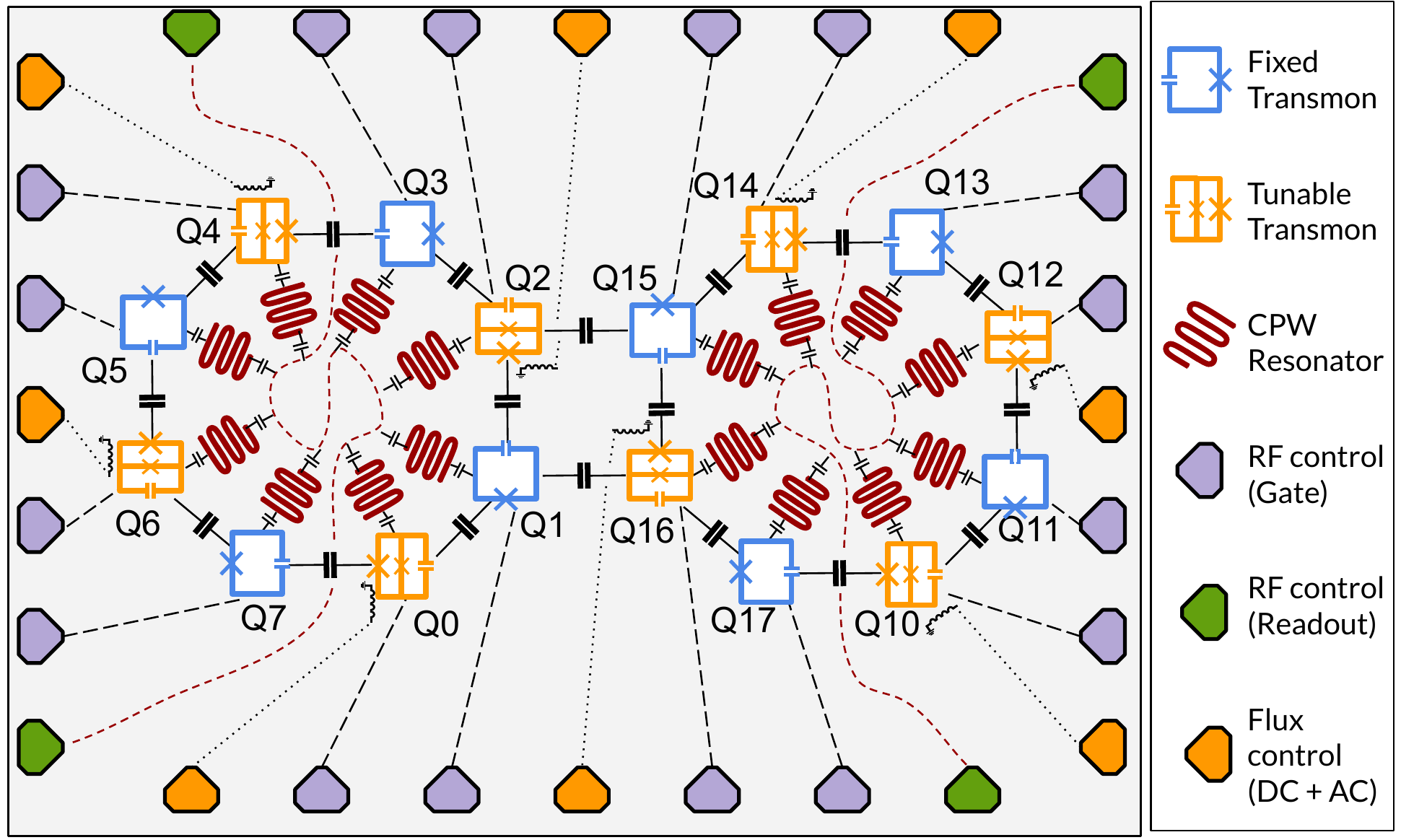}
    \caption{Circuit diagram of the device under test. Our planar architecture features eight frequency tunable transmons (orange) with a fixed capacitive coupling to eight fixed frequency transmons (blue). Readout employs eight-fold multiplexed quarter-wavelength coplanar waveguide resonators (red). Single qubit control is implemented by driving microwave pulses through each qubit’s microwave control line (purple). Frequency tunable transmons each have their own inductively coupled flux bias line that delivers alternating and direct currents which are combined outside the dilution refrigerator (orange). All control lines (dotted lines) are in the same plane as circuit elements and external control is brought in from contact pads at the edge of the chip, where individual aluminum wirebonds connect to a copper PCB.}
    \label{fig:chip}
\end{figure}

Our experimental device is a sixteen-qubit chip depicted schematically in Figure~\ref{fig:chip}, with relevant parameters given in Table~\ref{tab:device_table}. The chip's design comprises two octagonal groups of transmons, with eight-fold frequency multiplexing of readout within each octagon. Transmons in one octagon are numbered 0-7, and transmons in the other octagon are numbered 10-17. Each odd-numbered qubit is a fixed frequency single Josephson junction transmon, and each even-numbered qubit is a tunable transmon with an asymmetric DC SQUID configuration that allows its frequency spectrum to be tuned periodically by applying magnetic flux through the SQUID loop~\cite{PhysRevA.76.042319}. Each tunable transmon is coupled capacitively to its two or three nearest fixed-frequency neighbors. Charge control of each transmon is provided by a capacitively coupled voltage-control line in order to drive single qubit gates. Flux control is provided to each tunable transmon by a current-control line that is terminated near the SQUID loop in an inductive short to the common superconducting ground plane of the chip. The flux line is used to DC-bias the tunable qubits to set operating points, as well as to enact parametric gates with AC pulses~\cite{Didier:2018, caldwell2018parametrically, reagor2018demonstration}.

During the preparation of this manuscript, the device was available to users through Rigetti Computing's QCS platform as Aspen-4. The device was installed in a dilution refrigerator with filtering and signal delivery similar to that described in Ref.~\cite{Hong:2019}, and in particular the AC and DC flux-bias currents were combined at room temperature. The crosstalk between channels of the AWG was measured to be $\sim-80\,\mathrm{dB}$, or $0.01\%$, for frequencies less than 500~\,MHz and thus is not a significant contributing factor to observed on-chip crosstalk. Qubit 4 was omitted from measurements due to poor qubit performance.

\section{DC Crosstalk}\label{sec:dc-crosstalk}
In this section, we present two methods for measuring DC crosstalk. The first we call the \textit{resonator method}, which only involves measuring resonator frequencies. We refer to the second method as the \textit{qubit method}, which requires measuring the transmon frequencies.

The resonator method takes advantage of the coupling between each qubit and its respective readout resonator. We consider the undriven Hamiltonian describing two tunable transmons $A$ and $B$, coupled respectively to readout resonators $a$ and $b$. Transmon $i=A,B$ has first and second transition frequencies $\omega_i^{01}$ and $\omega_i^{12}$. The resonators have frequencies $\omega_a$ and $\omega_b$ and annihilation operators $\hat{a}$ and $\hat{b}$, and the transmons have truncated annihilation operators $\hat{q}$ and $\hat{p}$ describing only their first three energy eigenstates. We write the undriven Hamiltonian as
\begin{align}
\hat{H}/\hbar &\approx \omega_a \hat{a}^{\dagger} \hat{a} + \omega_A^{01} \hat{q}^{\dagger} \hat{q} + \eta_A \hat{q}^{\dagger} \hat{q}^{\dagger} \hat{q} \hat{q} / 2 \nonumber \\
& + \omega_b \hat{b}^{\dagger} \hat{b} + \omega_B^{01} \hat{p}^{\dagger} \hat{p} + \eta_B \hat{p}^{\dagger} \hat{p}^{\dagger} \hat{p} \hat{p} / 2 \nonumber \\
& + \chi_A \hat{a}^{\dagger} \hat{a} \hat{q}^{\dagger} \hat{q} + \chi_B \hat{b}^{\dagger} \hat{b} \hat{p}^{\dagger} \hat{p} 
+ \zeta \hat{q}^{\dagger} \hat{q} \hat{p}^{\dagger} \hat{p} 
\label{jc_ham}.
\end{align}
The dispersive shift $\zeta$ of the qubit-qubit interaction is negligible here, as the two tunable transmons are not intentionally coupled in the circuit. The dispersive shift $\chi_i=\frac{2g_i^2\eta_i}{\Delta_i (\Delta_i + \eta_i)}$, of transmon $i$ with its resonator, depends on the detuning $\Delta_i$ between the transmon and resonator, the bare coupling strength $g_i \sim 100$ MHz, and the transmon's anharmonicity $\eta_i \equiv \omega_i^{12} - \omega_i^{01}$. 
Taking the qubit to be in the ground state, we can absorb the $\chi_i$ terms of Eq.~\eqref{jc_ham} into the resonator terms to give dressed resonator frequencies which depend on $\Delta_i$. Each resonator therefore exhibits the same periodicity as its transmon with respect to flux, and the presence of an adversarial DC flux applied to qubit $B$ produces a shift in the phase of qubit A's resonator's periodic response. The magnitude of this phase shift is proportional to the flux crosstalk.

The advantage of only having to measure the resonators is that even on devices where low coherence times or extraneous resonances make accurate measurement of the qubit frequency difficult, the resonator method of measuring DC crosstalk can easily be performed. By measuring at least a full period of resonator A's tunability at several different values of qubit B's applied flux current, the dependence of the phase offset of resonator A on the flux current applied to qubit B can be determined. The flux offsets of adversarial qubit B are chosen to be $-\Phi_0$, 0, and $+\Phi_0$ so that qubit B's frequency is the same at all points and to confirm that changing the sign of the current changes the sign of the shift. 

\begin{figure}
    \includegraphics[width=\columnwidth]{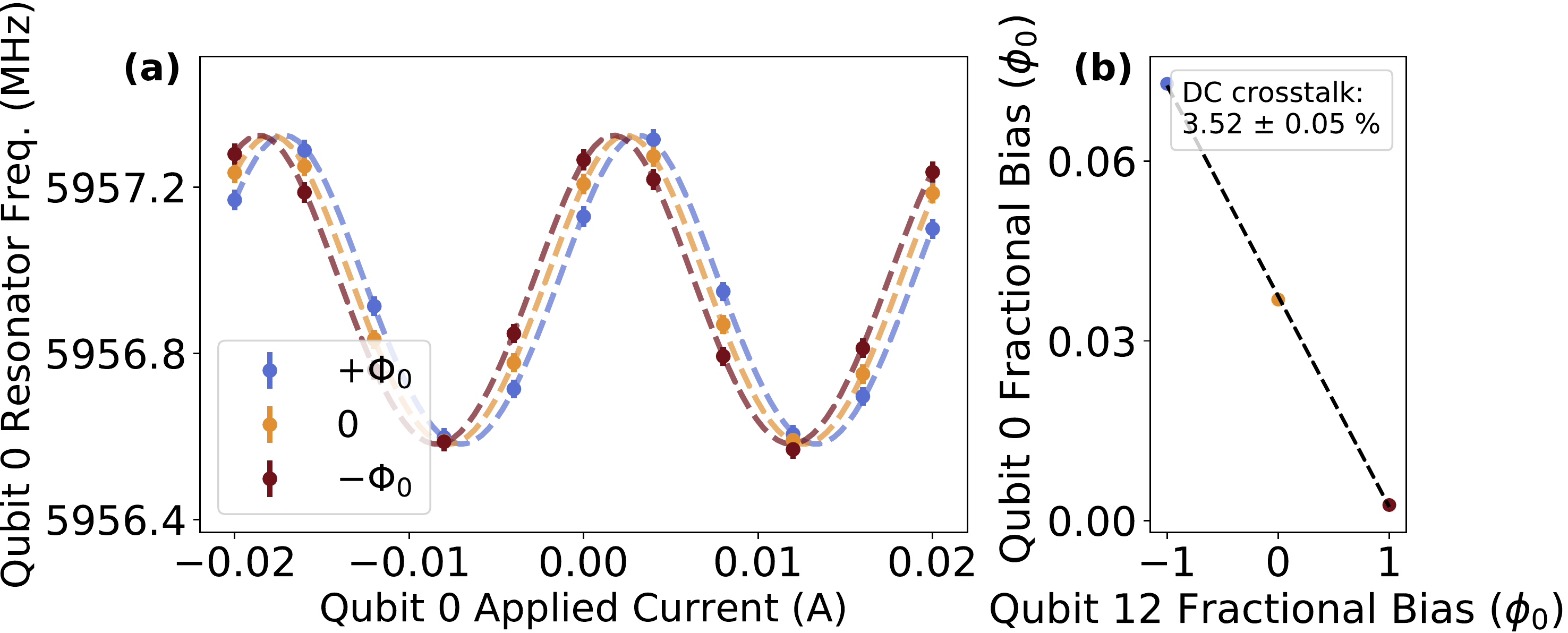}
    \caption{DC crosstalk via resonator spectroscopy. (a) Shows the frequency of Q0's resonator as a function of flux current applied at the top of the fridge to Q0's own flux line. The three different curves are taken at different values of Q12's flux current, as denoted in the plot legend. We fit the data to a model of a resonator coupled to a flux tunable transmon, keeping all free parameters fixed between the three fits except for the offset of the periodic curve. The shift in offset shows that Q0 is sensitive to flux applied to Q12's flux bias line. The error bars represent the $1\sigma$ error on the fit to the resonator frequency at each bias current. (b) Shows the fitted phase offsets of Q0's resonator plotted as a function of flux current applied to Q12 (in units of $\Phi_0$). The slope of the line gives the crosstalk between Q0 and Q12. The error bars are again $1\sigma$ errors extracted from the fit uncertainties in the phase offsets of the global fit used in (a).}
    \label{fig:dc_crosstalk_cav}
\end{figure}

Figure~\ref{fig:dc_crosstalk_cav} shows this method being used to measure the flux crosstalk between qubit A $= \mathrm{Q0}$ and qubit B $= \mathrm{Q12}$. The measured crosstalk of 3.5\% means that when flux current is applied to Q12 at DC, 3.5\% of the flux threading Q12's SQUID loop is also effectively threading Q0's SQUID loop.

\begin{figure}
    \includegraphics[width=\columnwidth]{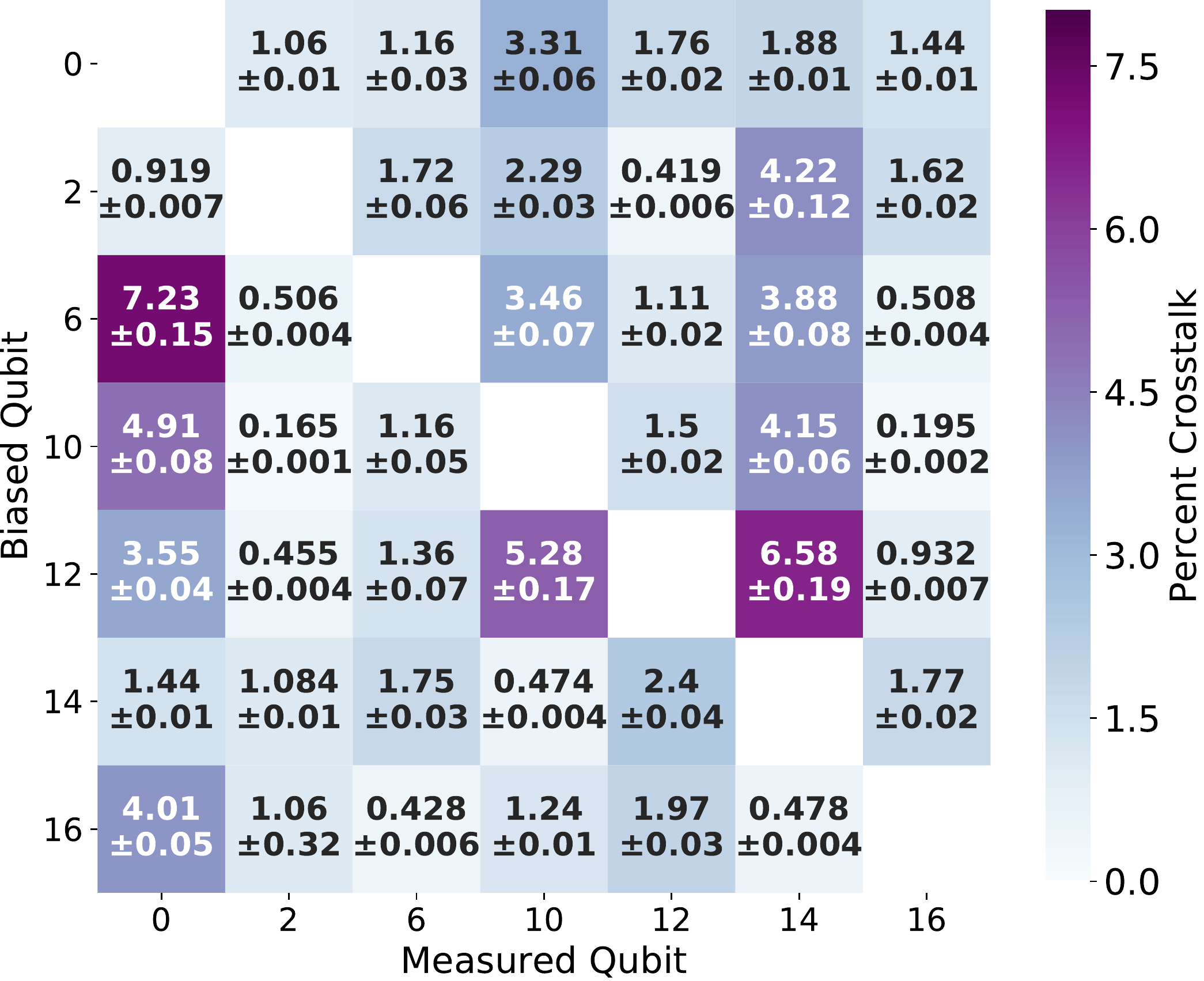}
    \caption{Matrix of DC crosstalk as measured by the qubit method, between all pairs of tunable qubits on the chip (excluding qubit 4). Errors are propagated from the $1\sigma$ errors on the linear fits of $df_A^{01}/d\Phi_A$ and $df_A^{01}/d\Phi_B$.}
    \label{fig:dc_crosstalk}
\end{figure}

Fitting the value of DC crosstalk using the resonator method involves accurately estimating the phase shift of three different periodic curves. For low values of crosstalk, it is difficult to do this with high accuracy. In order to get more precise measures of small values of DC flux crosstalk, the second method we present utilizes direct measurement of the qubits' frequency shifts. In this method we use measurements of the qubit frequency $f_A^{01}$ to determine flux crosstalk, via the relation
\begin{equation}
    \frac{d\Phi_A}{d\Phi_B} = \frac{df_A^{01}/d\Phi_B}{df_A^{01}/d\Phi_A}. \label{xtalk}
\end{equation}
To enhance the sensitivity of the measurement of $df_A^{01}/d\Phi_B$, we bias qubit A at $\Phi_A = \Phi_0/4$, near the point of maximum flux sensitivity. At this bias, qubit A's frequency response as a function of applied flux is roughly linear. We measure the two slopes $df_A^{01}/d\Phi_B$ and $df_A^{01}/d\Phi_A$ by measuring $f_A^{01}$ at three values of $\Phi_A$ about $\Phi_0/4$, and three values of $\Phi_B$ spaced in increments of $\pm\Phi_0$ to provide a large lever arm while holding constant the qubit-qubit interaction, as in the previous method. Each measurement of $f_A^{01}$ is performed using the Ramsey method. 

After performing this measurement pairwise across the chip, we can construct a matrix of DC flux crosstalk, as shown in Figure~\ref{fig:dc_crosstalk}. Looking at the entry for measuring Q0 and biasing Q12, one can see that the crosstalk of $3.55 \pm 0.04\%$ estimated from this method agrees with the $3.52 \pm 0.05\%$ estimated from the previous method. This showcases the consistency of the two methods for measuring DC crosstalk. A full comparison of the two methods across the whole matrix is given in Appendix~\ref{sup:xtalk compare}.

The qubit method achieves better sensitivity for extremely low values of crosstalk, while the resonator method does not require the excitation of any transmons in the circuit and is therefore robust to qubits that are difficult to measure. For our particular setup and sampling schemes, the sensitivity of the resonator method is $\delta \Phi \sim 100\,\mu\Phi_0$, given by the fit to the resonator response at each applied flux. With a Ramsey frequency sensitivity of $\delta f^{01} = 0.01\,\text{MHz}$, the qubit method provides a sensitivity of $\delta f^{01} / (df^{01}/d\Phi) = 5\,\mu\Phi_0$ on a tunable transmon with $df^{01}/d\Phi|_{\Phi_0/4} = 2000\,\text{MHz}/\Phi_0$. Both methods depend on a conversion between a flux quantum and units of the applied bias, which is readily obtained to a relative precision of less than 0.5\%. This can limit the relative uncertainty on the crosstalk value but is negligible in the limit of low crosstalk where sensitivity is crucial.

While the qubit method can achieve a finer sensitivity, it is vulnerable to distortions in the transmon's frequency response due to systematic effects such as: coherent interactions with the transmon's readout resonator, its neighboring transmons, or unwanted two-level systems in the device; severe losses of coherence near $\Phi_0/4$; or the nonlinearity of the transmon's frequency response to an adversarial flux of $\pm \Phi_0$ as the magnitude of crosstalk increases. The latter effect can be corrected with a sufficiently accurate mapping of the frequency response to flux, though that correction is not performed on the data presented here. Details of our implementation of each method are provided in Table~\ref{tab:dc-sensitivities}.

\begin{table}[]
    \begin{ruledtabular}
    \begin{tabular}{l | c c c}
        Method & DC, resonator & DC, qubit & AC \\
        \hline
        Sensitivity ($\mu\Phi_0$) & 110 & 5 & 20\\
        Shots per point & 200 & 300 & 500 \\
        Shot rate (kHz) & 50 & 10 & 10 \\
        Run time (secs) & 80 & 235 & 385 \\
        Excited transmon? & No & Yes & Yes
    \end{tabular}
    \end{ruledtabular}
    \caption{Sensitivities of each method for directly measuring flux crosstalk, as realized with the experimental parameters tabulated above. For each qubit, all measurements of the resonator and qubit responses were performed with the same signal power applied to the resonator. The run times quoted include latencies that are not captured by the shot rate and number of shots per point, and these latencies were not optimized for the purposes of this study. All sensitivities reflect fixed linear sampling schemes and could be enhanced by more optimal sampling strategies.}
    \label{tab:dc-sensitivities}
\end{table}

We observe that DC flux crosstalk is not symmetric between two qubits. For example, the entry for Q0 biasing Q16 is not equal to the entry for Q16 biasing Q0. The general lack of pattern or locality points to complex ground currents shaped by the chip layout. We have found these measures of DC crosstalk to be stable over time on the same chip, and relatively consistent among chips of the same design; however, crosstalk can vary dramatically between different chip designs.

\section{AC Crosstalk}\label{sec:ac-crosstalk}
Once qubits have been parked at a chosen operating point using a DC bias, we use AC flux pulses to operate two-qubit entangling gates. Our entangling gates are controlled by a pulsed AC signal $\Phiac \cos(\omega_p t + \theta)$, with $\omega_p/2\pi$ in the range of 50-500 MHz, and a modulation amplitude of $\Phiac$. These AC flux pulses produce a mean shift, $\bar{\Delta}=\bar{f}^{01}-f^{01}_\mathrm{max}$, of the qubit frequency from its parking frequency during the time the pulse is being played, where $\bar{f}^{01}$ is the qubit's average frequency under modulation, and $f^{01}_\mathrm{max}$ is the qubit's frequency when parked at 0 flux.  

\begin{figure}
    \includegraphics[width=\columnwidth]{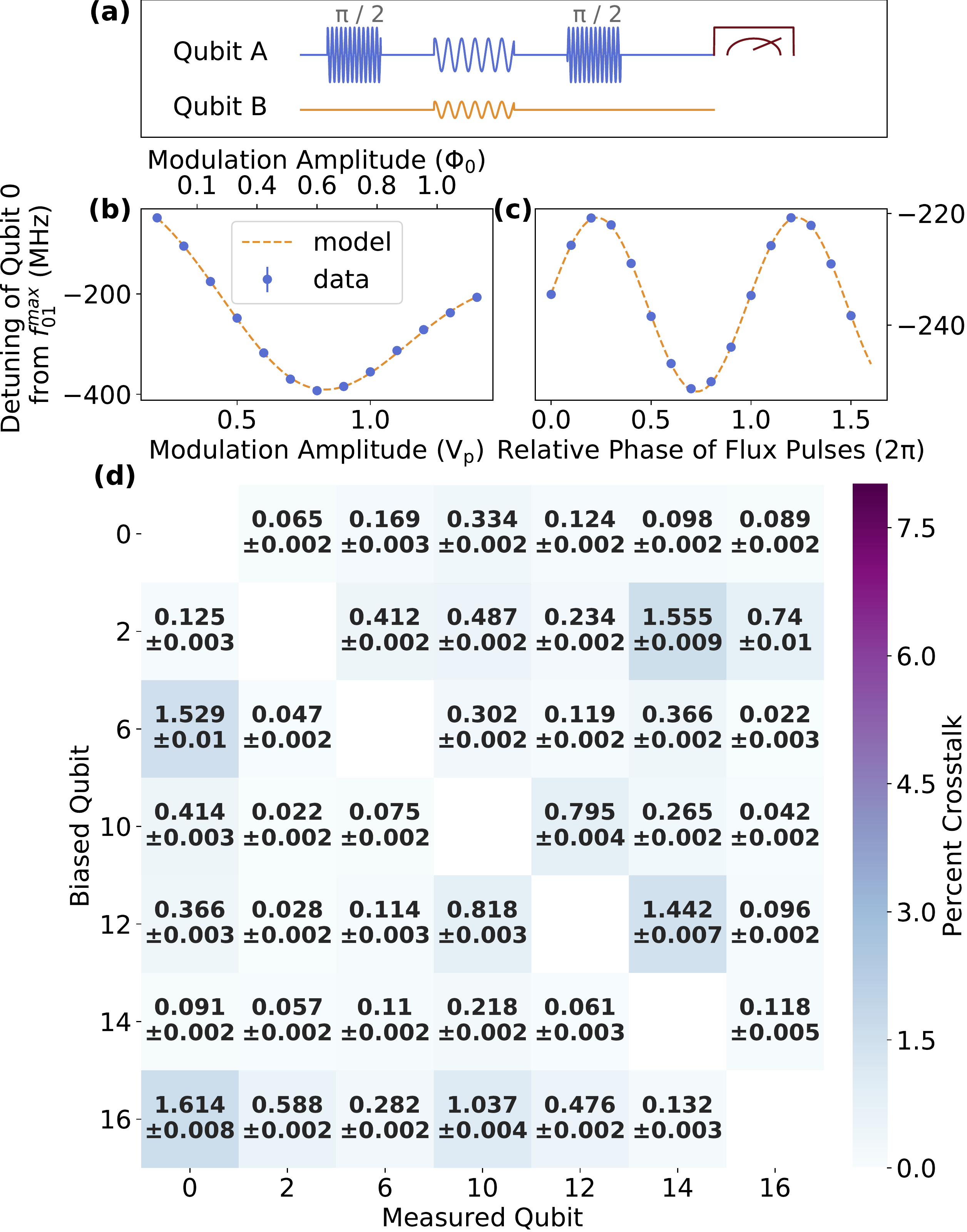}
    \caption{(a) Pulse sequence for measuring AC flux crosstalk. A Ramsey sequence is performed on the qubit to be measured (qubit A). During the wait time between the two $\pi/2$ pulses, flux pulses with the same frequency are played on qubit A and qubit B, causing qubit A to pick up a mean shift in qubit frequency, which is measured by the Ramsey experiment. Changing the relative phase of the two flux pulses causes their signals to constructively and destructively interfere, leading to a sinusoidal oscillation in the mean shift of qubit A's frequency. Qubit A is then measured. (b) Qubit 0 mean shift from $f^{01}_\mathrm{max}$ as a function of flux pulse amplitude, as measured by a Ramsey experiment with a single flux pulse played on qubit 0 during the wait period to detune the qubit. The data can be fit to Eq.~\eqref{mod_det_model}, allowing us to determine the conversion between flux pulse amplitude in volts and units of $\Phi_0$. (c) $\bar{\Delta}$ for Qubit 0 as a function of the relative phase between the two flux pulses (one played on Qubit 0, and one played on Qubit 6, as pictured in (a)). When the two pulses constructively interfere, the measured qubit is at its largest mean shift. When the two flux pulses destructively interfere, the measured qubit is at its smallest mean shift. The observed phase shift is due to some constant offset in the requested relative phase of the flux pulses. $1\sigma$ error bars are calculated based on fit uncertainties and plotted, but are smaller than the size of the data points. (d) AC flux crosstalk matrix for pairs of tunable qubits on the chip. All measurements were taken using flux pulses modulated at 200 MHz.}
    \label{fig:ac_crosstalk}
\end{figure}

When measuring AC flux crosstalk, we would like to again extract $d\Phiac_A/d\Phiac_B$, as in Eq.~\eqref{xtalk}, in order to easily compare with DC crosstalk. As in the qubit method of measuring DC crosstalk, wherein we park qubit A at its most DC-flux sensitive point, when measuring AC crosstalk we want to place qubit A at its most AC-flux sensitive point. We achieve this by sending a flux pulse to qubit A at an amplitude chosen to modulate the qubit to the linear regime of its $\bar{\Delta}$ vs. flux amplitude curve (see Figure~\ref{fig:ac_crosstalk}(b)). Once qubit A has been modulated to its most flux sensitive point, in order to study the effects of crosstalk, we send an adversarial flux pulse to qubit B at the same frequency as the flux pulse on qubit A, and measure qubit A's average frequency under modulation~\cite{Didier:2018} with a Ramsey experiment. This pulse sequence is shown schematically in Figure~\ref{fig:ac_crosstalk}(a). By sweeping the relative phases of these two flux pulses, we can observe increases and decreases of qubit A's mean shift from $f^{01}_{\mathrm{max}}$ as the pulses go in and out of phase with each other, resulting in a periodic response as shown in Figure~\ref{fig:ac_crosstalk}(c). From the amplitude of the response, we can fit $d\bar{f}^{01}_A/dV_B$.

In order to get the conversion between AC current at the top of the fridge and flux quanta at the bottom of the fridge for a given qubit, we characterize each qubit's frequency response as a result of sending a modulated flux pulse down that qubit's own flux line. Holding the modulation frequency constant to avoid frequency dependence in the signal chain transfer function, we sweep the modulation amplitude and measure $\bar{\Delta}$ for the qubit at each amplitude. The qubit's average frequency under modulation, $\bar{f}^{01}$, depends on the flux pulse amplitude $\Phiac$ as follows,
\begin{align}
\bar{\Delta} = \sum_{n=1}^\infty[\mathrm{J}_0(n2\pi\Phiac/\Phi_0)-1]\nu_n, \label{mod_det_model}
\end{align}
where the coefficients $\nu_n$ depend on the transmon's $E_{J1}$, $E_{J2}$ and $E_C$, and $\mathrm{J}_0$ is the $0^{\mathrm{th}}$ Bessel function of the first kind~\cite{Didier:2018}. Characteristic data from this measurement, as well as the fit of Eq.~\eqref{mod_det_model} to that data, are shown in Figure~\ref{fig:ac_crosstalk}(b).
One of the parameters of this model is the conversion rate between flux pulse amplitude and flux quanta, and thus the change in $\bar{\Delta}$ from the addition of an adversarial flux pulse can be converted to a change in flux through the qubit's SQUID loop using the parameters used to fit the model to the data. From these reference scans, we can extract $d\bar{f}^{01}_A/d\Phiac_A$ and $dV_B/d\Phiac_B$. We then have all the pieces we need to estimate $d\Phiac_A/d\Phiac_B$. Figure~\ref{fig:ac_crosstalk}(d) shows the result of this measurement between every viable pair of tunable qubits on the chip. With a Ramsey frequency sensitivity under modulation of $\delta\bar{f}^{01} = 0.05\,\text{MHz}$ and a slope $d\bar{f}^{01}/d\Phiac|_{0.35 \Phi_0} = 900\,\mathrm{MHz}$, this method provides a peak sensitivity (under constructive interference with the adversarial pulse) of $\delta\bar{f}^{01} / (d\bar{f}^{01}/d\Phiac) \sim 55\,\mu\Phi_0$. With the scan over the phase of the adversarial pulse as shown in Figure~\ref{fig:ac_crosstalk}(c), we reach a finer sensitivity of $\delta \Phiac \sim 20\,\mu\Phi_0$.

\begin{figure}
    \includegraphics[width=\columnwidth]{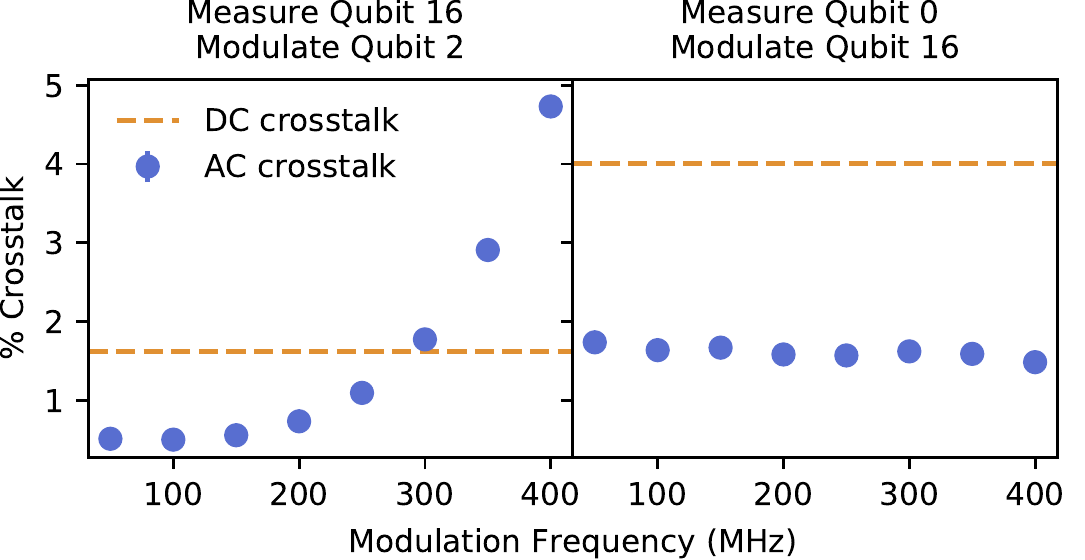}
    \caption{Examples of the variety of observed frequency dependence of AC flux crosstalk between pairs of tunable qubits. Horizontal dashed line shows measured DC crosstalk between that pair for comparison. (left) AC crosstalk between two qubits as a function of frequency with large response. (right) AC crosstalk between two qubits as a function of frequency with small response.}
    \label{fig:ac_crosstalk_freq}
\end{figure}

The measured AC crosstalk is typically smaller than the DC crosstalk between the same pair. It is natural to wonder exactly how crosstalk varies with frequency. We can measure AC flux crosstalk as a function of the frequency of the flux pulses. Figure~\ref{fig:ac_crosstalk_freq} shows the frequency dependence of AC crosstalk for two representative pairs on the chip. For the frequency dependence between all viable pairs on the chip, see Figure~\ref{fig:ac_xtalk_all_set}. For some pairs there is a large frequency dependence, while for other pairs the frequency dependence seems negligible. This frequency dependence suggests non-trivial transfer functions on the chip itself. We expect that AC and DC flux crosstalk converge at frequencies far below 50\,MHz; however due to bandwidth constraints of our control system, we were unable to explore the gap between DC and 50\,MHz. 

\section{Cross Tomography}\label{sec:cross-tomography}
In the previous section, it was shown that playing an AC flux pulse on qubit B can unintentionally detune qubit A. In this section, we examine the effects of that mean shift on qubit A's entangling gate fidelities. Our chosen entangling gate is a parametrically activated controlled-Z (CZ) gate, which has the unitary representation $U=\mathrm{diag}(1,1,1,-1)$. This gate is native to our architecture, as shown by the first harmonic terms of the Hamiltonian of a capacitively coupled fixed and tunable transmon under flux modulation in the interaction picture~\cite{Didier:2018}
\begin{align}
\hat{H}_\mathrm{int} &= g_{20}e^{i(2\omega_p-[\Delta+\eta_F])t}e^{i\beta_{20}}\,\ket{11}\bra{20} \\ 
& + g_{02}e^{i(2\omega_p-[\Delta-\eta_T])t}e^{i\beta_{02}}\,\ket{02}\bra{11}
+ \mathrm{H.c.}
\label{fullHint}
\end{align}
where $\Delta=\omega_{F_{01}}-\bar{\omega}_{T_{01}}$ is the detuning between the fixed-frequency transmon and the mean frequency of the tunable transmon, $\beta$ depends on the phase of the flux pulse, and we use the index ordering $\ket{\mathrm{Fixed},\mathrm{Tunable}}$. The CZ gate is realized by a $2\pi$ rotation $\ket{11} \rightarrow \{\ket{02}, \ket{20}\} \rightarrow -\ket{11}$ that is activated at a resonance frequency dependant on $\Delta$. The rotation rate is given by the renormalized coupling $g_{02}$ or $g_{20}$. Each CZ gate is activated by a flux pulse with a specific frequency and amplitude, calibrated such that the mean shift of qubit A and the frequency of the flux pulse combine to satisfy the resonance condition.

In the presence of flux crosstalk, simultaneous CZ gates suffer from the interference effect studied in section~\ref{sec:ac-crosstalk}. A CZ gate involving qubit A acts as an adversarial flux pulse to the flux pulse that activates a CZ involving qubit B, and vice versa. By shifting the mean frequency under modulation of each tunable transmon, the interference effect alters the entangling phases of both gates and the local phases of each tunable qubit. Both of these effects degrade the fidelities of the CZ gates. This fidelity degradation can be measured directly, giving us a computationally relevant measure of the effect of flux crosstalk.

\begin{figure}[h!]
    \includegraphics[width=\columnwidth]{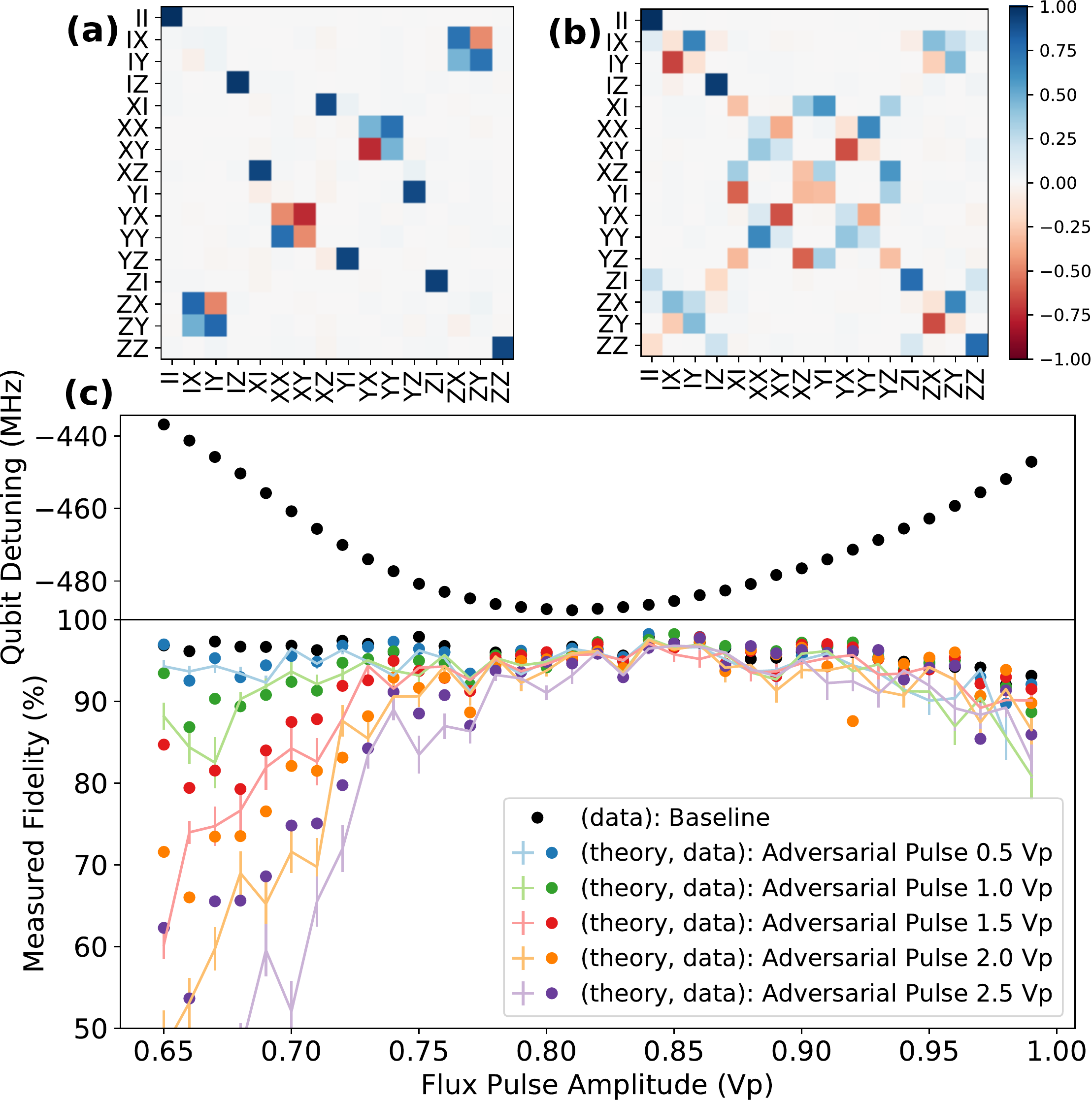}
    \caption{(a) Tomography of a CZ gate between qubits 1 and 2, with an adversarial CZ gate enacted between qubits 15 and 16. In this case, the measured process is equivalent to a CZ on qubits 1 and 2 along with additional single qubit Z rotations on those qubits (as indicated by the 2x2 blocks). (b) Tomography of a CZ gate between qubits 1 and 2, with an adversarial CZ gate enacted between qubits 1 and 16. In this case, the measured process cannot be described as a CZ plus single qubit Z rotations. (c) The top subplot shows qubit 2's $\bar{\Delta}$ as a function of flux pulse amplitude. The point of maximal $\bar{\Delta}$ is the AC sweet spot. The bottom subplot shows the measured fidelity of the CZ gate between qubits 1 and 2 as a function of the amplitude of the flux pulse used to perform the CZ gate. The black dots are the baseline fidelity measured with no adversarial pulse. The colored dots are the measured CZ fidelity while an adversarial pulse of the same frequency and duration is played on qubit 0, where each color represents a different amplitude (in Vp) of the adversarial pulse, used to simulate the effects of differing levels of flux crosstalk. The lines in lighter colors are the predicted gate fidelity for each adversarial flux pulse amplitude, as calculated by Eq.~\eqref{infid} using measured gate times and average frequency shifts due to crosstalk. The prediction is for the worst-case scenario where the phase of the adversarial pulse is such that the interference between the two flux pulses is maximal. When actually performing the simultaneous tomography measurement, the phase between the flux pulses is chosen to be zero at the controller; which, due to differing electrical delays, is not always the phase that leads to maximal interference. Discrepancies between the predicted and observed simultaneous gate fidelities may also be caused by potential additional dephasing effects due to the adversarial flux pulse.}
    \label{fig:cross_tomo}
\end{figure}

We measure this degradation in gate fidelity by performing quantum process tomography (QPT) on a CZ gate between two qubits while some other CZ is played simultaneously between another pair of qubits on the chip. QPT as a measure of fidelity has well known limitations~\cite{Shalm2005, Merkel2013}, however it is useful in providing diagnostics for the types of errors that we see. In some cases (Figure~\ref{fig:cross_tomo}(a)), the effect of the adversarial gate pulse is merely that the qubits involved in the gate under measure incur additional single qubit phase shifts (RZs). This can be seen by comparing the measured tomogram to the ideal process with additional RZs. By optimizing over the additional RZs to find the process that is the closest to the ideal one, we find that the baseline gate fidelity can often be almost entirely recovered by merely applying single qubit phase corrections. In these cases the crosstalk might be correctable with simple local frame updates. However, in other cases (Figure~\ref{fig:cross_tomo}(b)), more complex dynamics that arise from exciting other resonances, or more drastically shifting the qubit, result in measured processes that cannot be fully described by the ideal gate plus single qubit RZs.

In order to operate high-fidelity gates simultaneously, we would like to know whether there are certain operating points that are more resilient to crosstalk. In previous work, the concept of an ``AC sweet spot" for parametrically activated entangling gates has been discussed~\cite{Didier:2018}. This AC sweet spot is located at the point of maximal detuning of the tunable qubit from its parking frequency. At this point, the tunable qubit becomes first order insensitive to $1/f$ flux noise. Furthermore, the tunable qubit (and the gates enacted by modulating it) are less sensitive to crosstalk. This work can be expanded to predict the effects of crosstalk on CZ gate fidelity.
 
The leading order term contribution to a CZ's process infidelity, $r$, due to flux crosstalk (as derived in Appendix~\ref{sup:infid}) is equal to 
\begin{align}
r=(27\pi^2/20)(\delta\bar{f}_{01}\tau)^2 \label{infid}
\end{align}
where $\tau$ is the gate time and $\delta\bar{f}_{01}$ is the average frequency shift due to flux crosstalk.
For a given flux crosstalk $\XT\ll1$, a flux pulse of amplitude $\Phiac_B$ on qubit B (at the same frequency as a pulse on qubit A) shifts the effective modulation amplitude of the flux pulse on qubit A by $\delta\Phiac_A\approx\XT\Phiac_B\cos\theta$, where $\theta$ is the phase difference between the pulses. The shift of the average frequency is then $\delta\bar{f}_{01}=\frac{\partial\bar{f}_{01}}{\partial\Phiac_A}\delta\Phiac_A$, leading to a quadratic dependence of the infidelity with flux crosstalk. At the AC sweet spot, the slope $\frac{\partial\bar{f}_{01}}{\partial\Phiac_A}$ vanishes and the frequency shift is equal to $\delta\bar{f}_{01}=\frac{1}{2}\frac{\partial^2\bar{f}_{01}}{\partial\Phiac_A^2}\delta\Phiac_A^2$, leading to a quartic dependence of the infidelity with flux crosstalk. Thus, we expect gates operated at the AC sweet spot to be less susceptible to the deleterious effects of flux crosstalk. This calculation shows that it is possible to maintain simultaneous parametric gate fidelities above $99\,\%$ at the AC sweet spot for flux crosstalk $\XT<0.2\,\%$ in the worst case scenario of equal modulation frequencies.

In order to experimentally validate the connection between flux crosstalk and simultaneous two-qubit gate fidelities, we measure two-qubit gate fidelity between qubits 1 and 2 in the presence of an adversarial flux pulse at several different modulation amplitudes about the AC sweet spot. At each amplitude, the frequency and duration of the CZ gate is calibrated, and the gate fidelity is measured using quantum process tomography. Then, quantum process tomography is performed again on the qubits involved in the calibrated CZ gate (qubits 1 and 2); however, at the same time that the flux pulse for the CZ is being played, a flux pulse of the same frequency and duration is played on a different tunable qubit (qubit 0). This can be done at several different amplitudes of the adversarial pulse in order to simulate different levels of on-chip crosstalk. At each point, we also measure the change in qubit 2's frequency caused by the adversarial flux pulse on qubit 0, and use that value, along with the calibrated gate time, to calculate the expected drop in gate fidelity due to crosstalk as per Eq.~\eqref{infid}. The measured and theoretical fidelities are shown in Figure~\ref{fig:cross_tomo}(c) as dots and lines respectively. In general, the predicted fidelity tends to slightly overestimate the effects of crosstalk, as the change in qubit frequency due to the addition of an adversarial flux pulse used to calculate the effects of crosstalk is the worst-case scenario of the two flux pulses being completely in phase. However, the relative phase of the flux pulses used when measuring the gate fidelity was not always the worst-case phase.

It can be seen that the gate fidelity is more resilient close to the AC sweet spot, and at lower amplitudes of the adversarial pulse, i.e.~lower crosstalk, as predicted. It is also interesting to note that the resilience is not perfectly symmetric about the sweet spot as one might expect. This is due to the change in qubit 2's frequency being asymmetric about the sweet spot. A smaller change in $\bar{\Delta}$ at amplitudes higher than the sweet spot, and therefore also the resonance frequency of the gate, allows the maintenance of the baseline process fidelity.

At the sweet spot, the coherence limit of the gate fidelity in the absence of crosstalk is $97.94\%$, while the fidelity measured by QPT is $96.67\%$. The difference in expected and observed gate fidelity can be attributed to preparation error, which QPT is particularly susceptible to. The single qubit gate fidelities of qubits 1 and 2 measured in parallel are $99.82\%$ and $99.65\%$ respectively, thus we can expect to incur an error of $\sim 0.5\%$. That preparation error, along with natural fluctuations in qubit coherence times (and thus the coherence-limited fidelity), account for the observed infidelity of the entangling gate.

\section{Conclusion}
In summary, we have presented three different methods for directly measuring flux crosstalk, and demonstrated that crosstalk has a measurable impact on computationally relevant metrics such as simultaneous entangling gate fidelity. These measures of crosstalk can be predictive of simultaneous gate fidelities, and thus can be used as engineering milestones towards a fault-tolerant quantum computer. By crafting low-level measurements that are predictive of high-level behavior, we break the daunting problem of ensuring high-fidelity simultaneous entangling operations into more tractable chunks. We predict that a crosstalk level of $0.2\%$ or lower is required for simultaneous two-qubit gate operations at the sweet spot to maintain a fidelity of $99\,\%$. In order to achieve these levels of crosstalk we anticipate more deliberate and sophisticated routing of return currents on the chip will be required.

\begin{acknowledgments}

This work was funded by Rigetti \& Co Inc., dba Rigetti Computing. We thank the Rigetti quantum software team for providing tooling support, the Rigetti fabrication team for manufacturing the device, the Rigetti technical operations team for fridge maintenance, the Rigetti cryogenic hardware team for providing the chip packaging, and the Rigetti control systems and embedded software teams for creating the Rigetti AWG control system.

D.M.A, C.A.R, and S.A.C drafted the manuscript and designed the experiments.
D.M.A performed the experiments. 
N.D. provided theory to describe simultaneous gate fidelity degradation due to crosstalk. 
B.R.J. and C.A.R. organized the effort.
\end{acknowledgments}

\bibliography{references}

\appendix

\section{Comparison of Methods for Measuring DC Crosstalk}
\label{sup:xtalk compare}
\begin{figure}
    \includegraphics[width=\columnwidth]{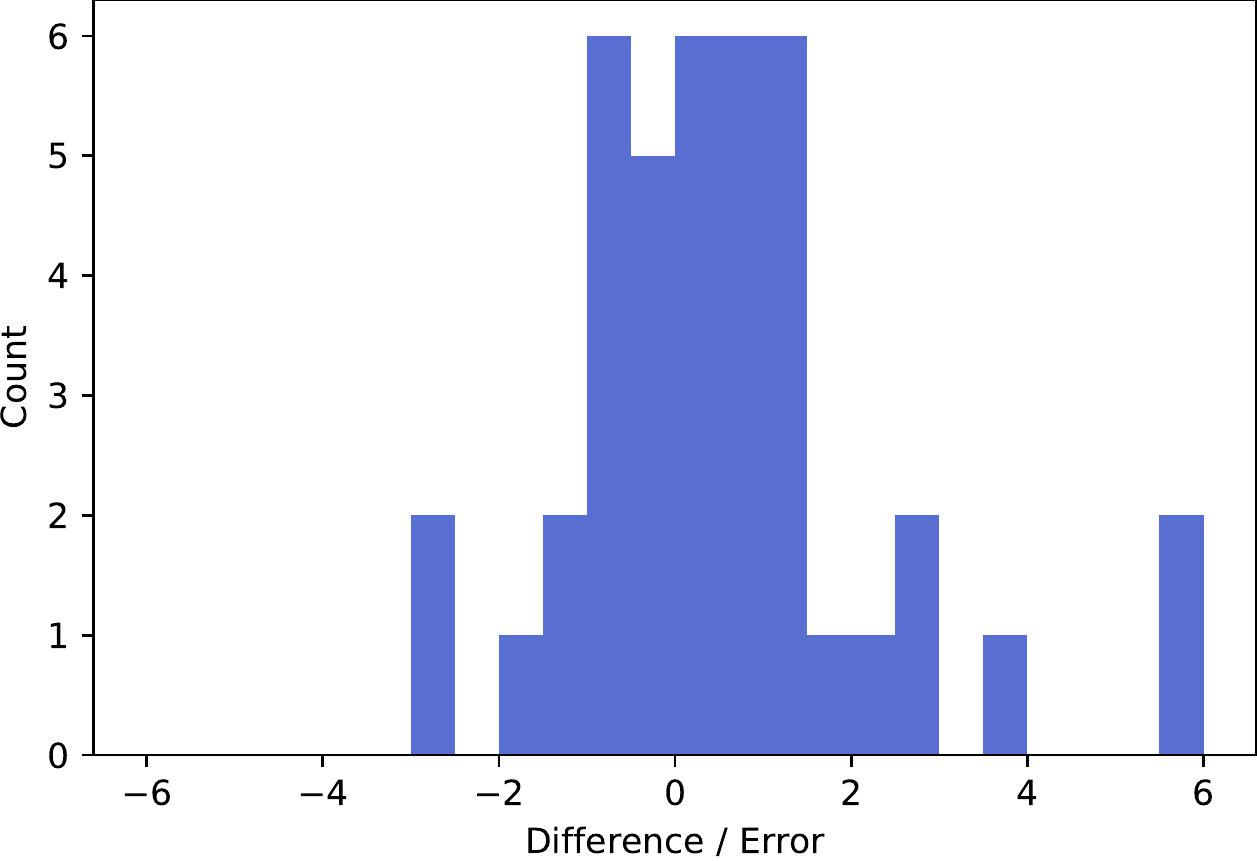}
    \caption{Histogram comparing the results of the two DC crosstalk measurement methods for each pair of tunable qubits on the chip. Plotted is the difference in reported crosstalk between the two methods, divided by the error bars for each method added in quadrature.}
    \label{fig:dc_crosstalk_compare}
\end{figure}
In Figure~\ref{fig:dc_crosstalk_compare}, we compare the two methods of measuring DC crosstalk for each pair on the chip. There are some pairs whose results are inconsistent between the two measurements; however, the difference in these cases is still below 0.1\%. Furthermore, the two measurement methods were taken 10 weeks apart, with a brief thermal cycle of the chip to 4K and back to base temperature during this time, which may have led to some of the inconsistencies.  

\section{Derivation of Infidelity Due to Crosstalk}
\label{sup:infid}
When an RF flux pulse $\Phi_A=\Phidc_A+\Phiac_A\cos(2\pi f_mt)$ is applied to tunable transmon A, and part of the flux current $\Phi_B=\Phidc_B+\Phiac_B\cos(2\pi f_mt+\theta_B)$ applied to qubit B at the same modulation frequency ($f_m$) biases qubit A, characterized by a crosstalk $\XT$, qubit A receives a total DC flux of $\Phidc_A(\XT)=\Phidc_A+\XT\Phidc_B$, and AC flux of amplitude
\begin{align}
\Phiac_A(\XT) 
& = \sqrt{\Phiac_A^2+\XT^2\Phiac_B^2+2\XT\Phiac_A\Phiac_B\cos\theta_B}\nonumber\\
& \approx \Phiac_A+\XT\Phiac_B\cos\theta_B,
\end{align}
in the limit of low crosstalk, $\XT\ll1$.

We consider the case where the adversarial flux pulse is played with a phase difference $\theta_B=0$ or $\pi$ and note that $\delta\Phiac=\Phiac_A(\XT)-\Phiac_A(0)\approx\pm\XT\Phiac_B$.
Changing the amplitude of the flux pulse seen by qubit A by $\delta\Phiac$ changes the transmon frequency under modulation.

When performing entangling gates, flux crosstalk changes the resonance condition and the effective coupling, making it necessary to re-calibrate the local Z rotations, the gate's modulation frequency, and the gate time to recover the fidelity. For a given mean shift $\bar{\Delta}$ of qubit A under modulation, and effective coupling without crosstalk $g_{\mathrm{eff}}$, the modulation frequency for the entangling gate, $f_m$, is proportional to $|\bar{\Delta}|/2$ and the gate time is $\tau=\pi/g_\mathrm{eff}$. In the presence of crosstalk, the change $\delta\Phiac$ in the modulation amplitude shifts the average qubit shift by $\delta\bar{\Delta}$ and the effective coupling between the qubits involved in the gate by $\delta g_\mathrm{eff}$.

To calculate the fidelity of a CZ gate in the presence of flux crosstalk, we use the interaction picture of the Hamiltonian 
$H_0=\omega_{01}(t)\ket{01}\bra{01}+\omega_{10}(t)\ket{10}\bra{10}+\omega_{11}(t)\ket{11}\bra{11}+[\omega_A(t)-\delta\bar{\Delta}]\ket{Q}\bra{Q}$,
where state $\ket{Q}=\ket{02}$ for CZ$_{02}$ and $\ket{Q}=\ket{20}$ for CZ$_{20}$.
Neglecting decoherence effects and the presence of harmonics, the Hamiltonian can be written as
\begin{align}
\hat{H}_\mathrm{int}&=\delta\bar{\Delta}\ket{Q}\bra{Q}+(g_{\mathrm{eff}}+\delta g_{\mathrm{eff}})[\ket{11}\bra{Q}+\ket{Q}\bra{11}],
\end{align}
Without crosstalk, $\delta\bar{\Delta}=\delta g_{\mathrm{eff}}=0$, and the gate fidelity is $100\,\%$.
The unitary operator from evolution under this Hamiltonian, projected in the logical basis, is
$U_\mathrm{int}(\tau)=\mathrm{diag}\{1,1,1,U_{11}(\tau)\}$
with
\begin{align}
U_{11}(\tau)&=-\left[\cos(\delta G\tau) -i\frac{\delta\bar{\Delta}}{2G}\sin(\delta G\tau)\right]e^{-i\frac{1}{2}\delta\bar{\Delta}\tau},
\end{align}
with $G=\sqrt{(g_{\mathrm{eff}}+\delta g_{\mathrm{eff}})^2+\frac{1}{4}\delta\bar{\Delta}^2}$ and $\delta G=G-g_{\mathrm{eff}}$.

The physics of the interaction picture is obtained by performing local Z rotations on the qubits to remove the dynamical phases, which are given by the time integral of the qubit frequency under modulation. 
For gate times much larger than the modulation period, this is given by the average qubit frequency.
These local operations are calibrated without crosstalk and the rotations applied to the tunable qubits are no longer correct in the presence of crosstalk, since crosstalk changes the average frequency by $\delta \bar{\omega}_{01}$. The final evolution operator is thus
$U(\tau)=\mathrm{diag}\{1,e^{-i\delta\bar{\omega}_{01}\tau},1,e^{-i\delta\bar{\omega}_{01}\tau}U_{11}(\tau)\}$.

The average infidelity is equal to 
$r=(d^2-|\mathrm{tr}\{U_\mathrm{CZ}^\dag U(\tau)\}|^2)/(d^2+d)$ with $d=4$.
At leading order in $\delta\bar{\omega}_{01}$, $\delta\bar{\Delta}$, $\delta g_{\mathrm{eff}}$, we find

\begin{align}
r=\tfrac{1}{5}(\delta\bar{\omega}_{01}\tau-\tfrac{1}{4}\delta\bar{\Delta}\tau)^2 + \tfrac{1}{40}(\delta\bar{\Delta}\tau)^2 + \tfrac{1}{5}(\delta g_\mathrm{eff} \tau)^2.
\end{align}

The term with $\delta g_{\mathrm{eff}}$ can be suppressed by tuning the gate time,
the imperfection $\delta\bar{\omega}_{01}$ is removed by applying the relevant local Z rotation, 
the term $\delta\bar{\Delta}$ is removed by considering a $\mathrm{CPhase}(\pi-\frac{1}{2}\delta\bar{\Delta})$ instead of $\mathrm{CZ}=\mathrm{CPhase}(\pi)$.

The average change in the shift, $\delta\bar{\Delta}$, depends on the type of CZ that is implemented.
It is equal to $-\delta\bar{\omega}_{01}$ for $\mathrm{CZ}_{02}$ and to $\delta\bar{\omega}_{01}$ for $\mathrm{CZ}_{20}$.
This gives rise to the following infidelities,

\begin{align}
r_{02}& = \frac{27}{80}(\delta\bar{\omega}_{01}\tau)^2 + \frac{1}{5}(\delta g_{\mathrm{eff}} \tau)^2, \label{infidelity}\\
r_{20}& = \frac{11}{80}(\delta\bar{\omega}_{01}\tau)^2 + \frac{1}{5}(\delta g_{\mathrm{eff}} \tau)^2.
\end{align}

The average frequency shift and the change of effective coupling are found from their slope with respect to modulation amplitude, 
$\delta\bar{\omega}_{01}=\frac{\partial\bar{\omega}_{01}}{\partial\Phiac}\delta\Phiac$ and 
$\delta g_{\mathrm{eff}}=\frac{\partial g_{\mathrm{eff}}}{\partial\Phiac}\delta\Phiac$.
This holds away from the AC flux sweet spot where, by definition, the slope of the average frequency is zero. At the AC flux sweet spot, the average frequency shift is calculated from the curvature, $\delta\bar{\omega}_{01}=\frac{1}{2}\frac{\partial^2\bar{\omega}_{01}}{\partial\Phiac^2}\delta\Phiac^2$.
This is also the case for the effective coupling when the device is designed to have the maximum effective coupling at the AC sweet spot.

For the CZ$_{02}$ gate, assuming that $\delta g_{\mathrm{eff}}$ is negligible and using the fact that $\omega = 2\pi f$, we recover Eq.~\eqref{infid}.

\section{Device Parameters}

\begin{table*}
\begin{ruledtabular}
\begin{tabular}{r | c c c c}
& Readout Resonator (GHz) & Qubit $f_{01}$ (GHz) & Qubit Anharmonicity (MHz) &  Qubit-Resonator $\chi$ (MHz) \\
\hline
Q0 & 5.957 & 4.678 & -186 & -0.59 \\
Q1 & 5.657 & 3.821 & -206 & -1.29 \\
Q2 & 5.913 & 4.759 & -187 & -0.76 \\
Q3 & 5.695 & 3.767 & -235 & -1.30 \\
Q5 & 5.739 & 3.919 & -203 & -1.41 \\
Q6 & 5.836 & 4.639 & -189 & -0.75 \\
Q7 & 5.781 & 3.880 & -204 & -1.41 \\
Q10 & 5.950 & 4.766 & -161 & -0.75 \\
Q11 & 5.657 & 3.573 & -186 & -1.08 \\
Q12 & 5.914 & 4.849 & -183 & -0.95 \\
Q13 & 5.696 & 3.646 & -208 & -1.21 \\
Q14 & 5.872 & 4.767 & -188 & -0.78 \\
Q15 & 5.737 & 3.753 & -206 & -1.17 \\
Q16 & 5.831 & 4.480 & -191 & -0.54 \\
Q17 & 5.775 & 3.533 & -212 & -1.05 \\
\end{tabular}
\end{ruledtabular}
\caption{Relevant parameters for the device under test. We list here the resonator readout frequencies, the qubit transition frequencies, the qubit anharmonicities, and the coupling between each qubit and its resonator. For even numbered qubits, which are frequency-tunable transmons, quoted values of qubit frequency, anharmonicity, and qubit-resonator $\chi$ were measured with the qubit parked at 0 flux.}
\label{tab:device_table}
\end{table*}

\section{AC Crosstalk as Function of Frequency}
\label{app:xtalk_as_freq}

\begin{figure*}
    \includegraphics[width=\textwidth]{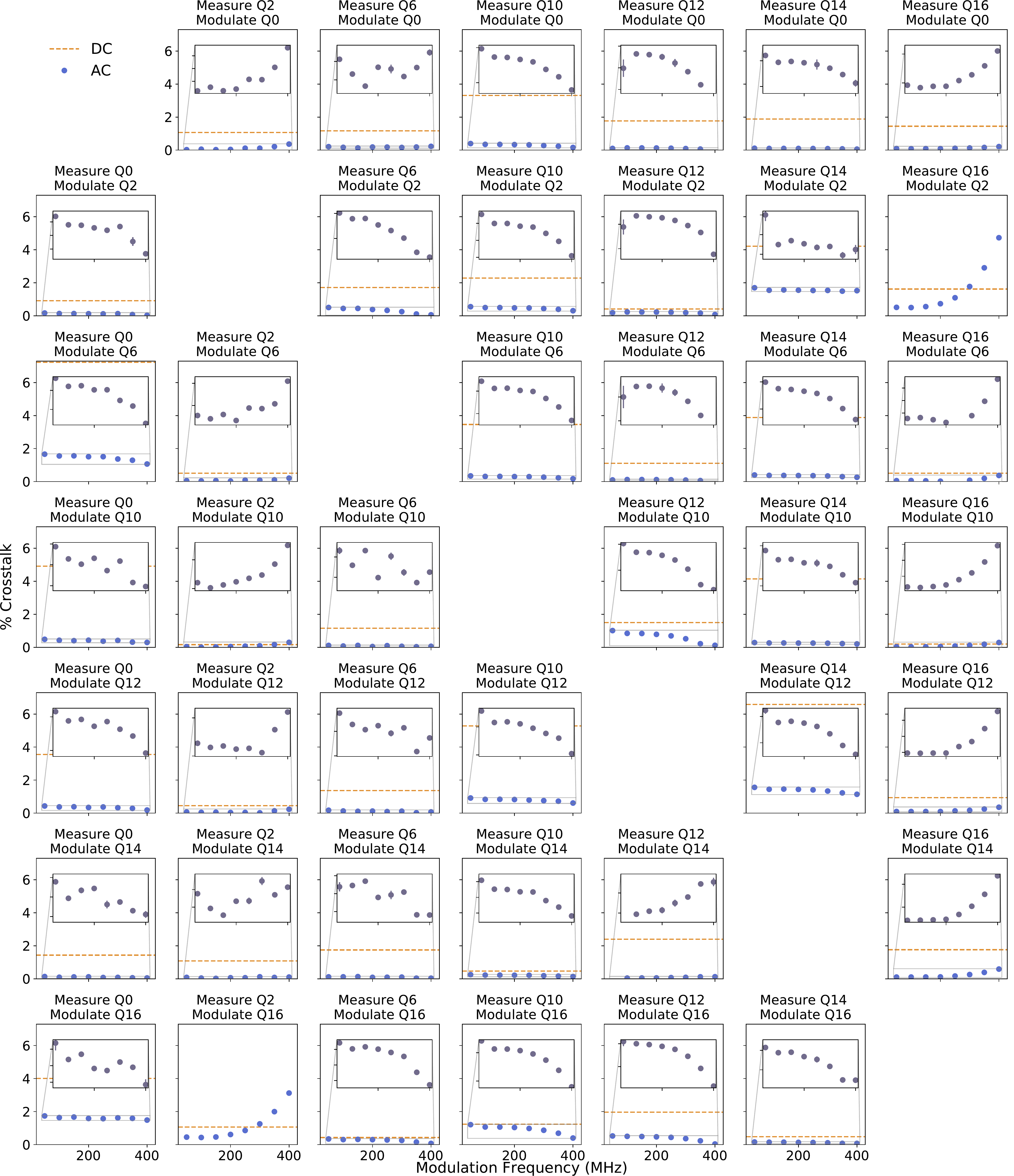}
    \caption{AC flux crosstalk between pairs of tunable qubits across the chip. AC crosstalk values are marked as blue dots, while dashed yellow lines show measured DC crosstalk for comparison. All y-axes have the same scale for easy cross-pair comparison; while the insets show the smaller-scale structure for each pair.}
    \label{fig:ac_xtalk_all_set}
\end{figure*}

\end{document}